\documentclass[aps,pra,twocolumn,showpacs]{revtex4}

\usepackage{graphicx}
\usepackage{latexsym}
\usepackage{amssymb}
\usepackage{color}
\usepackage{bm}
\usepackage{amsmath}
\usepackage{subfigure}

\begin{document}

\title{Phase diagram of the Bose Kondo-Hubbard model}

\author{Michael Foss-Feig}
 \affiliation{
 JILA, National Institute of Standards and Technology, and University of Colorado, Boulder, Colorado 80309, USA}
\author{Ana Maria Rey}
\affiliation{
JILA, National Institute of Standards and Technology, and University of Colorado, Boulder, Colorado 80309, USA}

\begin{abstract}
We study a bosonic version of the Kondo lattice model with an on-site
repulsion in the conduction band, implemented with alkali
atoms in two bands of an optical lattice.  Using both weak and
strong-coupling perturbation theory, we find that at unit filling of
the conduction bosons the superfluid to
Mott insulator transition should be accompanied by a magnetic
transition from a ferromagnet (in the superfluid) to a paramagnet (in
the Mott insulator).  Furthermore, an analytic treatment of Gutzwiller
mean-field theory reveals that quantum spin fluctuations induced by the Kondo
exchange cause the otherwise continuous superfluid to Mott-insulator
phase transition to be first order.  We show that lattice separability
imposes a serious constraint on proposals to exploit excited bands for
quantum simulations, and discuss a way to overcome this constraint in
the context of our model by using an experimentally realized non-separable lattice.  A method to
probe the first-order nature of the transition based on collapses and
revivals of the matter-wave field is also discussed.
\end{abstract}
\date{\today}
\pacs{03.75.Ss, 37.10.Jk, 67.85.--d, 71.27.+a, 75.30.Hx}

\maketitle

\setlength{\parskip}{0pt plus .5mm minus .2 mm}
\setlength{\abovecaptionskip}{3 mm}
\setlength{\belowcaptionskip}{-1 mm}

\section{Introduction}
Ultracold atoms in optical lattices have been used to
simulate a variety of condensed matter Hamiltonians \cite{blochRMP},
with eminent successes including the simulation of both the Bose \cite{greiner} and
Fermi \cite{schneider,jordens} \emph{single-band}
Hubbard models.  More recently, progress in controlling and
stabilizing atoms in the excited bands of
an optical lattice
\cite{mullerbloch,wirth,anderlini,PhysRevA.73.020702} has given rise
to the exciting possibility of simulating \emph{multi-band} condensed matter Hamiltonians, which involve a
nontrivial interplay of spin, charge, and orbital degrees
of freedom.  These achievements have precipitated a variety of
theoretical investigations into the new physics made possible by
the orbital degrees of freedom in an optical lattice \cite{larson,zhou,PhysRevLett.101.125301,PhysRevLett.101.186807}.

The Kondo lattice model (KLM), in which tightly bound
electrons act as spinful scattering centers for electrons in a
conduction band \cite{Tsunetsugu:1997p597}, is a typical example of the type of model one would like to
simulate.  In the KLM, the orbital degree of freedom gives rise to a
rich phase diagram that includes, e.g., magnetically ordered states, a heavy
Fermi liquid, and unconventional superconductors.  In this manuscript
we will revisit a version of the KLM first
proposed in Ref. \cite{duan}, in which the electrons are replaced by
spin-$\frac{1}{2}$ bosons, with the spin degree of freedom encoded in
two hyperfine states of an alkali atom.  Our primary new finding is that, for any small but nonzero Kondo
coupling, the typically continuous superfluid (SF) to Mott insulator (MI)
phase transition becomes first-order.  The qualitative difference
between the pure Hubbard and Kondo-Hubbard model, even at arbitrarily
weak Kondo coupling, is reminiscent of similar results for the Fermi
Kondo-Hubbard model obtained in
Ref. \cite{assaad}.  That the inclusion of small inter-band
interactions (which are often relevant in real materials) can
have such a dramatic effect on the Bose Hubbard phase diagram
underscores the importance of generalizing optical lattice simulations
to include orbital degrees of freedom.

The structure of the manuscript is as follows.  The experimental
implementation of the Bose Kondo-Hubbard model is discussed in
Sec. \ref{themodel}.  This section deviates from the original proposal
\cite{duan} in that we emphasize the essential role of lattice
non-separability in obtaining the model in more than one spatial dimension.  We begin our investigation of the
model's phase diagram in Sec. \ref{effectivespinmodels}, where we derive
effective spin Hamiltonians valid in the weak coupling
(Sec. \ref{effectivespinmodelsWC}) and strong coupling
(Sec. \ref{effectivespinmodelsSC}) limits in order to understand the magnetic
properties of the SF and MI phases.  We
then employ an analytic treatment of Gutzwiller mean-field theory
(MFT) in Sec. \ref{meanfieldtheory} to map out the ground
state phase diagram.  At mean-field level, one can observe the
interplay of two competing tendencies: Superfluidity of the
conduction bosons tends to spontaneously break $\mathrm{SU}(2)$ symmetry, whereas the Kondo interaction
prefers a rotationally symmetric ground state composed of localized singlets.  While the rigidity of the superfluid protects it from
magnetic fluctuations, we will see that these fluctuations give rise to a metastable MI of Kondo
singlets, causing the MI to SF transition to become first-order.  Such
a first-order transition has been realized---and confirmed by Quantum Monte Carlo---in the spin-$\frac{1}{2}$ boson
model of Ref. \cite{batrouni}.  In Sec. \ref{opticallattice} we consider
experimental details related to dynamically maintaining a two-band
model.  In particular we discuss the implementation of the model, and
the relevant parameter regimes, using the non-separable
lattice of Ref. \cite{sebbystrabley}. In Sec. \ref{detection} we discuss the preparation of the
unit-filled MI phase.  We then suggest the possibility of experimentally observing the first-order
transition by ramping down the lattice to enter the SF regime;
due to the discontinuous nature of the phase transition, even an
arbitrarily slow ramp should excite collapses and revivals of the superfluid order
parameter.  In Sec. \ref{conclusion} we summarize our findings and
their relevance.

\section{The model\label{themodel}}
Everything that follows assumes a two dimensional (2D) optical lattice $V(\bm{r})$ populated by bosonic
alkali atoms with mass $m$ and $s$-wave
scattering length $a_s$.  We assume that all atoms are in the $F=1$ hyperfine
manifold with $m_F=\pm1$.  In $^{87}$Rb, the similarity of scattering
lengths for total spin $F=0$ and $F=2$ collisions, together with the
ability to offset the $m_F=0$ state via the quadratic Zeeman
effect \cite{anderlini,HoandZhang}, strongly suppresses spin changing
collisions into the $m_F=0$ state.  This justifies considering only
two internal states for times long after the initial preparation,
and we label these two internal (spin) degrees of freedom by
$\sigma=\uparrow,\downarrow$.  At temperatures sufficiently low that
$s$-wave scattering dominates,
and neglecting nearest neighbor interactions, the Hamiltonian
describing this system is
\begin{widetext}
\begin{equation}
\label{HAM}
\mathcal{H}=\sum_{ij\alpha\sigma}J_{\alpha ij}\alpha^{\dagger}_{i\sigma}\alpha^{}_{j\sigma}+\sum_{j\alpha}\frac{U_{\alpha}}{2}\hat{n}_{j\alpha}(\hat{n}_{j\alpha}-1)+\!\!\sum_{j,\alpha>\beta}\!\!V_{\alpha\beta}\hat{n}_{j\alpha}\hat{n}_{j\beta}+\!\!\!\!\sum_{j,\alpha>\beta,\sigma\sigma'}\!\!\!\!V_{\alpha\beta}\alpha^{\dagger}_{j\sigma}\beta^{\dagger}_{j\sigma'}\alpha^{}_{j\sigma'}\beta^{}_{j\sigma}+\!\!\!\!\!\!\!\!\!\!\sum_{j,\{\alpha,\beta\}\neq\{\gamma,\delta\}}\!\!\!\!\!\!\!\!W_{\alpha\beta\gamma\delta}\alpha^{\dagger}_{j\sigma}\beta^{\dagger}_{j\sigma'}\gamma^{}_{j\sigma'}\delta^{}_{j\sigma}
\end{equation}
\end{widetext}
In Eq. (\ref{HAM}), $\alpha^{\dagger}_{j\sigma}$ creates a spin-$\sigma$ boson in a
Wannier orbital $w_{\alpha}(\bm{r}-\bm{r}_j)$ of the
$\alpha^{\mathrm{th}}$ band, located on the lattice site centered at
$\bm{r}_j$.  The density operator for bosons
on site $j$ in the $\alpha^{\mathrm{th}}$ band is
$\hat{n}_{j\alpha}=\sum_{\sigma}\alpha^{\dagger}_{j\sigma}\alpha^{}_{j\sigma}$,
and the various parameters are given by
\begin{eqnarray}
J_{\alpha ij}&=&\int\;d^3\bm{r}w_{\alpha}(\bm{r}-\bm{r}_i)\left[V(\bm{r})-\frac{\hbar^2\nabla^2}{2m}\right]w_{\alpha}(\bm{r}-\bm{r}_j) \nonumber \\
U_{\alpha}&=&\int\;d^3\bm{r}w_{\alpha}(\bm{r})^4 \nonumber \\
V_{\alpha\beta}&=&\int\;d^3\bm{r}w_{\alpha}(\bm{r})^2w_{\beta}(\bm{r})^2 \nonumber \\
W_{\alpha\beta\gamma\delta}&=&\int\;d^3\bm{r}w_{\alpha}(\bm{r})w_{\beta}(\bm{r}) w_{\gamma}(\bm{r})w_{\delta}(\bm{r}).
\end{eqnarray}
The notation $\{\alpha,\beta\}\neq\{\gamma,\delta\}$ means that these
two sets of indices cannot contain identical elements, and is
introduced so that the final term in $\mathcal{H}$ contains all (and
only) scattering processes that
change the band populations on a given site.  The hoppings
$J_{\alpha ij}$ are so far unspecified; we will only require, since we
want a 2D model, that all such hoppings emanating from one site to its
nearest neighbors are similar in magnitude.
 To set the overall scale of kinetic energy we define the effective hopping
strength $J_{\alpha}=\frac{1}{4}\sum_{i}|J_{\alpha ij}|$ \footnote{The factor of $\frac{1}{4}$ is so that $J_{\alpha}$ agrees with the
  usual hopping integral in the case of nearest neighbor
  hopping on an isotropic square lattice.}.

We further restrict our attention to an initial
state with one atom per site in the lowest vibrational band ($b$ band,
or localized band) of the lattice, and an average of $n$ atoms per site in a single excited
band ($a$ band, or conduction band). The assumption that these conditions will be maintained dynamically,
a necessary condition for obtaining a bosonic Kondo model, amounts to
disregarding all of the $W_{\alpha\beta\gamma\delta}$ terms in Eq. (\ref{HAM}).  In
previous works \cite{girvin,mullerbloch,PhysRevA.73.020702}, such an approximation has been justified largely by
the restricted density of states in an optical lattice.  Very roughly,
the argument is that when interactions are small compared to typical band gaps, and if the bands themselves
are narrow, then band changing collisions tend to be off resonant and
strongly suppressed.  Here we point out a notable exception to the validity of this
reasoning, which to our knowledge has not been previously described in
the literature.  This exception affects the implementation of the
Kondo-Hubbard model, and also of the multi-flavor models described in
Ref. \cite{girvin}.

\begin{figure}[!h]
\centering
\subfiguretopcaptrue

\subfigure[][]{
\includegraphics[width=4.0 cm]{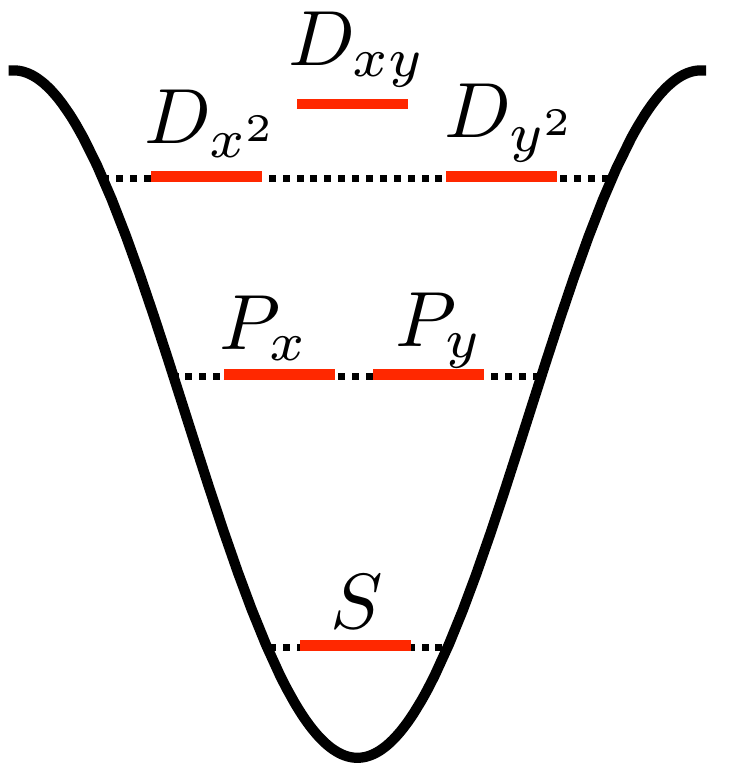}
\label{SPA}}
\subfigure[][]{
\includegraphics[width=4.0 cm]{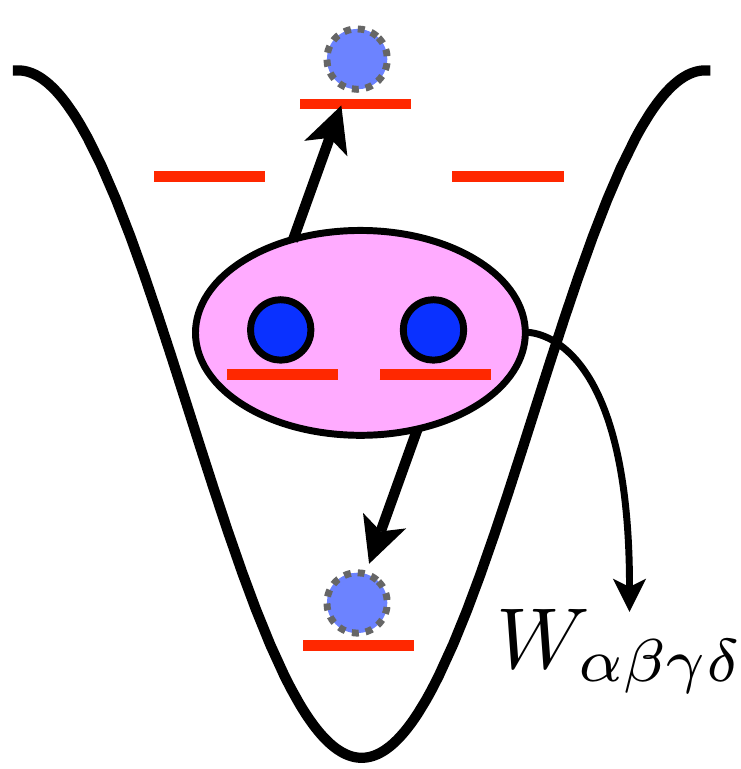}
\label{SPB}}
\caption{(Color online) In \subref{SPA} we plot an energy level diagram for states
  $|00\rangle$ ($S$ band), $|10\rangle$ ($P_x$ band), $|01\rangle$
  ($P_y$ band), $|20\rangle$ ($D_{x^2}$ band), $|02\rangle$ ($D_{y^2}$
  band), and $|11\rangle$ ($D_{xy}$ band).  The resonant
  scattering process (due to the $W_{\alpha\beta\gamma\delta}$ terms
  in $\mathcal{H}$) that transfers atoms from the initially
  populated $S$ and $D_{xy}$ bands into the $P_x$ and $P_y$ bands is shown in \subref{SPB}.}
\label{SeparableProblem}
\end{figure}

For a lattice that can be written $V(x,y)=V_x(x)+V_y(y)$, a Wannier orbital in the $a$ band can be written
$w_{\alpha\beta}(x,y)=w_{\alpha}(x)w_{\beta}(y)$, whereas a Wannier
orbital in the $b$ band is $w_{00}(x,y)=w_{0}(x)w_{0}(y)$.  To
have a 2D model, it must be true that hopping in both the $x$ and $y$
directions is greater in the conduction band than in the localized
band, and hence $\alpha,\beta>0$.  Because of the separability of the
non-interacting part of the Hamiltonian in the $x$-$y$ basis, the state
$|00\rangle\otimes|\alpha\beta\rangle$ [describing a single site with one atom in
$w_{00}(x,y)$ and one atom in $w_{\alpha\beta}(x,y)$] is, even in the
presence of interactions, \emph{exactly} degenerate with the state
$|0\beta\rangle\otimes|\alpha0\rangle$.  Furthermore, these states are
connected by the interaction term $W$ with a matrix element equal to the
exchange integral $V$ between the conduction and localized orbitals.  The situation is depicted
graphically in Fig. \ref{SeparableProblem} for the case where
$\alpha=\beta=1$, and the states $|00\rangle$, $|\alpha0\rangle$,
$|0\beta\rangle$, and $|\alpha\beta\rangle$ belong to the $S$, $P_x$,
$P_y$, and $D_{xy}$ bands respectively.  The $P_x$ and $P_y$
bands will be populated via collisions on a timescale $\tau=2\pi/V$; this is unacceptable,
since $\tau$ will turn out to be the timescale for singlet formation,
and one can't expect to see Kondo like physics if the approximations
yielding the model break down so quickly.  The way around
this problem is to use an optical lattice potential which cannot be
separated in cartesian coordinates.  In section \ref{detection} we
will give an example of an existing non-separable lattice which is
favorable for avoiding this problem, but for the time being we
continue with the approximation that the $W_{\alpha\beta\gamma\delta}$
can be ignored.

If the atoms in the $b$ band are deep in the unit-filled MI regime,
and we drop terms that are therefore constant, then $\mathcal{H}$ can
be reduced to a bosonic Kondo-Hubbard model \cite{duan}
\begin{eqnarray}
\label{Hamiltonian}
\mathcal{H}_{\mathrm{K}}&=&-\sum_{ij\sigma}J^{}_{ij}a^{\dagger}_{i\sigma}a^{}_{j\sigma}+\frac{U}{2}\!\sum_{j}\hat{n}_{ja}(\hat{n}_{ja}-1)\nonumber\\
&+&2V\!\sum_j\bm{S}_{ja}\cdot\bm{S}_{jb}-\mu\sum_{j}\hat{n}_{ja}.
\end{eqnarray}
In Eq. (\ref{Hamiltonian}) a chemical potential has been included to facilitate the
forthcoming mean-field treatment, and the spin
operators are defined by $\bm{S}_{j\alpha}\equiv\frac{1}{2}\sum_{\sigma\sigma'}\alpha^{\dagger}_{j\sigma}\bm{\tau}^{}_{\sigma\sigma'}\alpha^{}_{j\sigma'}$,
with $\bm{\tau}$ being a vector whose components are the Pauli
matrices.  In order to avoid a clutter of indices, we have defined
$V\equiv V_{ab}$, $J_{ij}\equiv J_{ija}$ (and $J\equiv J_a$), and
$U\equiv U_a$. Because $V$ has the same sign as $a_s$, the
Kondo interaction is antiferromagnetic (AFM) for repulsive
interactions, which is a manifestation of Hund's rule adapted for
bosons: Antisymmetrization of the spin wave function for two
identical bosons requires the antisymmetrization of their spatial
wavefunction, thus keeping them apart and lowering the energy cost of repulsive interactions.

\section{Effective spin models\label{effectivespinmodels}}
In the presence of the Kondo term, one still expects the conduction
bosons to undergo a MI to SF phase transition as the ratio $U/J$ is varied.  However, the magnetic properties of
these phases will be heavily influenced.  The goal of this section is
to study magnetic ordering in the SF and MI phases by
deriving effective spin Hamiltonians that are accurate in either the
weak coupling or strong coupling limits.  The results
presented are meant to reinforce and complement the
mean-field picture that will subsequently be developed.

\subsection{Weak coupling\label{effectivespinmodelsWC}}
In the fermionic KLM at $V=0$, the localized spins are decoupled from
the conduction electrons, giving rise to a large spin degeneracy in
the ground state manifold.  The lifting of this degeneracy is the result of virtual excitations of
the conduction electrons from below to above the Fermi-surface, which
mediate long ranged couplings between the localized spins known as
the Rudderman-Kittel-Kasuya-Yosida (RKKY) interaction \cite{randk,kasuya,yosida}.  In dimensions
greater than one, the RKKY interaction is believed to stabilize long-range magnetic order
in the KLM ground state \cite{kasuya}.  The primary difference in the bosonic model is
that the $V=0$ spin degeneracy extends to the conduction band, where the
ferromagnetic superfluid can point in any direction.  Surprisingly, this
additional freedom actually simplifies the situation; the degeneracy will now be
lifted at first (rather than second) order in $V$, and at this order the effective
Hamiltonian can be solved exactly.

Although we are not considering the presence of a confining potential
in this manuscript, in order to preserve the greatest generality in the
ensuing discussion we will state our results in terms of arbitrary
single particle eigenstates $\psi_r(j)$ and eigenvalues $\epsilon_r$
for a lattice plus trap.  At $V,U=0$, the ground
state is formed by putting all conduction bosons in the single
particle wavefunction $\psi_0(j)$ and has a degeneracy of
$(N_a+1)\times2^{\mathcal{N}}$, where $\mathcal{N}$ is the number of
lattice sites (also the number of $b$ atoms), and $N_a$ is the number
of $a$ atoms.  It is useful to define the functions $\mathcal{G}^r_{jl}=\psi^{*}_{r}(j)\psi^{}_{r}(l)$, in
terms of which the spin density operator for a conduction boson in the single
particle groundstate is 
\begin{equation}
\bm{S}^0_a=\frac{1}{2}\sum_{jl}\mathcal{G}^{0}_{jl}\sum_{\sigma\sigma'}a^{\dagger}_{j\sigma}\bm{\tau}_{\sigma\sigma'}a^{}_{l\sigma'}.
\end{equation}
Projection of $\mathcal{H}_{\mathrm{K}}$ onto the
degenerate groundstate manifold yields an effective weak coupling
Hamiltonian that is first order in the Kondo coupling $V$
\begin{equation}
\label{csm}
\mathcal{H}^{(1)}_{\mathrm{wc}}=2V\bm{S}^0_a\cdot\sum_{j}\mathcal{G}_{jj}^0\bm{S}^{}_{jb}.
\end{equation}
The first order energy shift due to $U$ has been dropped because
it does not lift the spin degeneracy.  Equation (\ref{csm}) describes the so-called \emph{central spin model}, which is familiar in the context of electron spin decoherence
in quantum dots \cite{PhysRevLett.88.186802}.  In the application to quantum dots, the model
describes coupling between the spin of an electron in the dot and the nuclear spins of the atomic
lattice in which the dot sits, with a coupling function
determined by the square of the electron's wavefunction.  In the present case, the
condensate atoms are Schwinger bosons representing the central spin $\bm{S}^0_a$, which couples to the mutually non-interacting localized spins with a
coupling function given by the square of the condensate wavefunction.
The central spin model has been studied extensively in the literature, and exact solutions have been
obtained by Bethe ansatz \cite{gaudin}.  However, for a translationally invariant system $\mathcal{H}^{(1)}_{\mathrm{wc}}$ can easily
be diagonalized by rewriting it in terms of the conserved quantities
$s_b$, $s_a$, and $s$, where $s_{\alpha}$ is the total spin quantum
number of $\bm{S}_{\alpha}=\sum_{j}\bm{S}_{j\alpha}$, and $s$ is the total spin quantum number of the combine
spin $\bm{S}_a+\bm{S}_b$.  The ground state
is formed when the localized spins align fully
ferromagnetically ($s_b=\mathcal{N}/2$), and then their total spin $\bm{S}_b$ couples as antiferromagnetically as possible to the condensate spin $\bm{S}_a$ ($s=\frac{\mathcal{N}}{2}|n-1|$).

Second-order perturbation theory in the Kondo exchange generates a correction to the weak coupling
Hamiltonian
\begin{eqnarray}
\label{wceh}
\mathcal{H}_{\mathrm{wc}}^{(2)}&=&-n\mathcal{N} V^2\sum_{j,l}\mathcal{R}_{jl}\bm{S}_{jb}\cdot\bm{S}_{lb} \nonumber \\
&+& 2V^2\bm{S}_a\cdot\sum_{j}\mathcal{R}_{jj}\bm{S}_{jb},
\end{eqnarray}
where
$\mathcal{R}_{jl}=\sum_{r}\mathcal{G}^r_{jl}\mathcal{G}^0_{lj}(\varepsilon_r)^{-1}$
[see Appendix \ref{wdderivation} for a derivation of Eq. (\ref{wceh})].
The first term in Eq. \ref{wceh} is the bosonic analog of the RKKY interaction for
fermions, and has the same physical origin; a spin at site $j$
scatters an $a$ atom out of the condensate,
which then re-enters the condensate upon scattering off the spin at
site $l$, and in this way the two spins talk to each other.  The second
term, which renormalizes the spin couplings of
$\mathcal{H}_{\mathrm{wc}}^{(1)}$, reflects the ability
of the scattered boson to return to the condensate with its spin
flipped.  Such a term is absent for fermions because Pauli exclusion
principle prevents a conduction fermion from returning to the Fermi sea
with its spin flipped\footnote{This is strictly true in the
  thermodynamic limit.  For a finite fermionic system an intensive
  central spin like term exists whenever the number of conduction
  fermions is odd.}.  Unlike the RKKY term this correction is local
in space, a property that can be attributed to the cancellation of time reversed
scattering processes involving more than one localized spin (the sign
for such a scattering process depends on whether the spin flip occurs
when the conduction atom is exiting or returning to the condensate).

Although the first term has the same physical origin as RKKY for
fermions, an important difference arrises due to the structure
of the coupling function.  In the fermionic case,
the Fermi surface introduces a length scale $k_F^{-1}$, at which the
long-range coupling oscillates.  For noninteracting bosons there is no
such length scale, the coupling function $\mathcal{R}$ does not
oscillate, and one can show that it is strictly positive at any
finite separation for a translationally invariant system in the thermodynamic limit.  Thus any
groundstate of $\mathcal{H}^{(1)}_{\mathrm{wc}}$ automatically
minimizes $\mathcal{H}^{(2)}_{\mathrm{wc}}$, and the polarization of
the localized spins persists to second order in V (note also that the
renormalization of the central spin coupling constants by the second
term in $\mathcal{H}^{(2)}_{\mathrm{wc}}$ is toward larger
positive values).  It should be noted, however, that $\mathcal{N}\mathcal{R}_{jj}$ diverges
logarithmically with the system size in 2D, suggesting that the energy cannot actually be expanded in
powers of $V$.  Nevertheless, the existence of exclusively
ferromagnetic (FM) terms in the first two orders of perturbation theory
strongly suggests a FM ground state at weak coupling.  We note that at
unit filling, the true groundstate of $\mathcal{H}_{\mathrm{wc}}$ is a singlet
($s=0$), but nevertheless the superfluid and the localized spins are
aligned ferromagnetically within themselves
($s_a=s_b=\frac{\mathcal{N}}{2}$).  That $s_b=\frac{\mathcal{N}}{2}$
means that all off-diagonal elements of the correlation function
\begin{equation}
\chi_{jl}=\langle\bm{S}_{jb}\cdot\bm{S}_{lb}\rangle
\end{equation}
obtain the maximum value of $\frac{1}{4}$.  Hence when we say that the $n=1$ superfluid phase is FM,
we mean that ferromagnetism exists independently within the superfluid
and within the localized spins.

\subsection{Strong coupling\label{effectivespinmodelsSC}}

We will take the strong coupling limit to be defined by $U\gg J$, for
any $V$, and for simplicity we will restrict our discussion to the case of
commensurate filling in the conduction band.  However, we will see that a similar limit
arrises for $V\gg J$ and any $U$ when the $a$ atoms are at
unit filling.  For integer $n\geq1$ the ground state is a MI with $n$ conduction bosons per
site.  The eigenstates on a single site follow from the addition of angular
momenta $\bm{S}_{ja}+\bm{S}_{jb}\equiv\bm{S}_j$, and
therefore have total spin quantum number $s_j^{\pm}=(n\pm1)/2$.  Because the
interaction is AFM, the eigenstate with lower total spin
is the ground state, and the MI phase with density $n$ must have total
spin $(n-1)/2$ at each site.  If on a single site we label the state
by its total spin quantum number $s_j$ and the $z$ projection of total
spin $s_j^z$, then the $n$ filling MI at $J=0$ is given by
\begin{equation}
|\mathrm{MI}_n\rangle=\bigotimes_j |(n-1)/2,s_j^z\rangle,
\end{equation}
and it has energy per site $E_n=\frac{U}{2}n(n-1)-\frac{V}{2}(2+n)$.
For $n=1$ each site contains a singlet $|0,0\rangle_j$, spin excitations
are gapped ($\Delta_{\mathrm{s}}=2V$), and the ground state is PM.  We also note
that for unit filling the charge gap is $\Delta_{\mathrm{c}}=U+V$, and does not
vanish as $U\rightarrow0$.  Hence the unit-filled MI exists whenever either $U$ or $V$ is large compared to $J$.

States with $n\geq2$ can be obtained from the singlet by repeated application
of the $a$ atom creation operators:
\begin{equation}
|s^-_j,s_j^z\rangle\propto(a^{\dagger}_{j\uparrow})^{s^-_j+s_j^z}(a^{\dagger}_{j\downarrow})^{s^-_j-s_j^z}|0,0\rangle_j.
\end{equation}
It is interesting to note that for a MI phase with $n\geq2$ the charge
gap is given by $\Delta_{\mathrm{c}}=U$, and has no dependence on $V$.  Therefore
the limit $V\gg J$ alone does not yield a MI in this case.  Because
the states $|s^{-}_j,s^z_j\rangle$ are degenerate ($s_j^z=-s^-_j,...,s^-_j$), it is
possible to derive an effective super-exchange Hamiltonian between the
total spins on neighboring sites by perturbation theory in the
hoppings $J_{ij}$.  The calculation---though complicated by the existence of virtual
excited states with two possible total spin quantum numbers---is straightforward, and yields an effective
Hamiltonian that only contains the \emph{total spin operators} $\bm{S}_j$
\begin{equation}
\label{sch}
\mathcal{H}_{\mathrm{sc}}=-g(n)\sum_{ij}\frac{J_{ij}^2}{U}\bm{S}_i\cdot\bm{S}_j.
\end{equation}
In Eq. (\ref{sch}),
\begin{equation}
g(n)=\frac{4n+8}{(n+1)^2}\left(\frac{V(n+1)^2+U(n+2)^2}{Vn+U}\right)
\end{equation}
is a strictly positive, density dependent coupling
constant, and we have dropped an overall density dependent energy shift.  The
ground state is therefore FM with total spin
$\frac{\mathcal{N}}{2}(n-1)$, and, unlike for the $n=1$ MI, there is no spin gap.  Notice that this result for the total
spin also holds for the weak coupling case, although the nature of the FM state
is completely different in the two limits.  While the inter-site spin
correlations between the localized spins are maximized at weak coupling, in the strong-coupling
limit we find the diminished correlation
$\chi_{jl}=\frac{1}{4}(\frac{n-1}{n+1})^2$, which vanishes when
$n=1$.

\section{Mean Field Theory\label{meanfieldtheory}}
As was shown by a numerical analysis of
the Gutzwiller variational ansatz in Ref. \cite{duan}, in the presence of
the localized spins the $a$ atoms continue to exhibit a MI to SF phase
transition at commensurate filling.  Here we
adopt an alternate but equivalent description of the Gutzwiller
variational ansatz as a site-decoupled mean-field theory (MFT) \cite{fisher}, and
obtain an analytical description of the phase transition.  The
motivation for this mean-field treatment is that (1) it
semi-quantitatively describes the MI to SF phase transition of bosons in $D>1$ and (2)
the new features of the present model---the Kondo term---are entirely local, and thus are
included without further approximation.

\begin{figure*}[!t]
\centering
\includegraphics[width=17.5 cm]{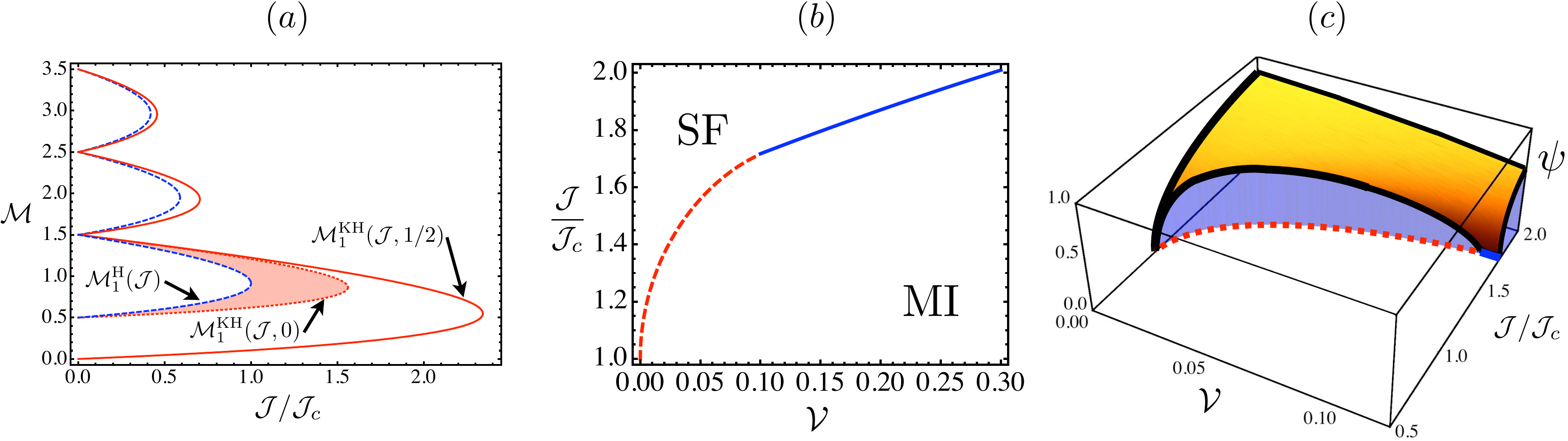}
\caption{(Color online) (a) Mott lobes in the $\mathcal{M}-\mathcal{J}$ plane, with
  $\mathcal{J}$ plotted in units of $\mathcal{J}_c=(3+\sqrt{8})^{-1}$,
  which is where the single band mean-field transition occurs at unit
  filling.  The red solid lines depict the boundaries of Mott lobes for
  $V=U/2$, given by $\mathcal{M}_n^{\mathrm{KH}}(\mathcal{J},1/2)$.  
  The blue dashed lines are the boundaries of Mott lobes for the
  single band boson Hubbard model, denoted
$\mathcal{M}_n^{\mathrm{H}}(\mathcal{J})$.  These have been shifted up by $\mathcal{V}=1/2$
along the $\mathcal{M}$-axis in order to more easily compare their widths with
those of the Kondo-Hubbard lobes.  The red dotted line is the boundary
$\mathcal{M}_1^{\mathrm{KH}}(\mathcal{J},0)$, which does not agree
with $\mathcal{M}_n^{\mathrm{H}}(\mathcal{J})$.  In Fig. \ref{JVphasediagram}(b) we plot the phase boundary of
the $n=1$ MI in the $\mathcal{J}$-$\mathcal{V}$ plane.  The solid blue
line demarks a continuos transition, while the red dotted line demarks
a first-order transition.  Fig. \ref{JVphasediagram}(c) shows the
superfluid order parameter $\psi$ in the $\mathcal{J}$-$\mathcal{V}$
plane, showing the discontinuity along the first-order portion of the
phase transition.}
\label{JVphasediagram}
\end{figure*}

The starting point of Gutzwiller mean-field theory is the decoupling
of the kinetic energy term
\begin{equation}
a^{\dagger}_{i\sigma}a^{}_{j\sigma}\rightarrow\langle a^{\dagger}_{i\sigma}\rangle
  a^{}_{j\sigma}+a^{\dagger}_{i\sigma}\langle
  a^{}_{j\sigma}\rangle-\langle a^{\dagger}_{i\sigma}\rangle\langle a^{}_{j\sigma}\rangle
\end{equation}
Assuming that the mean-field order parameter $\psi_{\sigma}=\langle
a_{j\sigma}\rangle$ is translationally invariant, and defining
$\mathcal{J}=4J/U$, $\mathcal{V}=V/U$, and $\mathcal{M}=\mu/U$, we can rewrite
the exact Hamiltonian $\mathcal{H}_{\mathrm{K}}$ as a sum over identical Hamiltonians defined at
each site (any one of which we call the mean-field Hamiltonian):
\begin{eqnarray}
\mathcal{H}_{\mathrm{GA}}&=&-\mathcal{J}\sum_{\sigma}(\psi^{}_{\sigma}a^{\dagger}_{\sigma}+\psi^*_{\sigma}a^{}_{\sigma})+\mathcal{J}\sum_{\sigma}\psi^*_{\sigma}\psi^{}_{\sigma}\nonumber
\\
&+&\frac{1}{2}\hat{n}_a(\hat{n}_a-1)+2\mathcal{V}\bm{S}_a\cdot\bm{S}_b-\mathcal{M}\hat{n}_a.
\label{HGA}
\end{eqnarray}
The mean-field ground state is obtained by minimizing
$E_{\mathrm{GA}}(\psi_{\sigma})\equiv\langle\mathcal{H}_{\mathrm{GA}}\rangle$
with respect to $\psi_{\sigma}$.  The order parameter describes the formation of superfluid coherence;
if $\psi_{\sigma}$ is finite, then by virtue of having to point somewhere it describes a state that
spontaneously breaks SU(2) symmetry, as the superfluid should.  On the
other hand, $\psi_{\sigma}=0$ is the signature of a Mott insulator phase, in
which number fluctuations are suppressed and phase coherence
vanishes.  Because $\mathcal{H}_{\mathrm{GA}}$ is invariant under the U(1)
gauge transformations $a_{\sigma},\psi_{\sigma}\rightarrow
e^{i\theta_{\sigma}}a_{\sigma},e^{i\theta_{\sigma}}\psi_{\sigma}$, we are justified to choose both components
of our spinor order parameter to be real.  Physically, this choice
amounts to confining
any superfluid that arrises to live in the $x-z$ plane, since the $y$
component of the (vector) superfluid density
\begin{equation}
\bm{\rho}=\sum_{\sigma\sigma'}\psi^{*}_{\sigma}\bm{\tau}_{\sigma\sigma'}\psi_{\sigma'}
\end{equation}
vanishes when $\psi^{*}_{\sigma}=\psi_{\sigma}$.  It
should be understood in what follows that this apparent broken
symmetry is an artifact maintained for
simplicity.

The mean-field Hamiltonian can easily be solved numerically by
truncating the single-site Hilbert space to contain no more than some
finite number of $a$ atoms.  However,
much insight can be gained by proceeding as far as possible
analytically.  If one makes the assumption that the SF to MI
transition is continuous, then the mean-field phase boundary is described exactly
by perturbation theory in $\psi_{\sigma}$.  We want to emphasize from
the beginning that this assumption will later be shown to
break down in certain parameter regimes.  Nevertheless, with this
assumption in mind one writes the mean-field
energy as
\begin{equation}
E_{\mathrm{GA}}=\mathcal{A}(\mathcal{J},\mathcal{V},\mathcal{M})+\mathcal{B}(\mathcal{J},\mathcal{V},\mathcal{M})\psi^2+\mathcal{O}(\psi^4),
\end{equation}
where $\psi^2\equiv\sum_{\sigma}\psi_{\sigma}^2$ is the superfluid
density, and the phase boundary is determined by the
condition $\mathcal{B}=0$.  That $E_{\mathrm{GA}}$ can be written as
an even function of $\psi$ follows from the symmetries of
$\mathcal{H}_{\mathrm{GA}}$.  For a Mott lobe of filling $n$, the groundstate at finite
$\mathcal{V}$ is in general degenerate (see Sec. \ref{effectivespinmodelsSC}), and
$\mathcal{B}$ must be found by calculating the
lowest eigenvalue of the effective Hamiltonian
\begin{equation}
\label{Heff1}
\mathcal{H}_{\mathrm{eff}}=\mathcal{J}^2\sum_{n\sigma\sigma'}\psi_{\sigma}\psi_{\sigma'}\frac{(a^{\dagger}_{\sigma}+a^{}_{\sigma})|n\rangle\langle
  n|(a^{\dagger}_{\sigma'}+a^{}_{\sigma'})}{E_0-E_{n}}.
\end{equation}

When $n=1$ the MI ground state is unique (the
singlet), and while degenerate perturbation theory is unnecessary it
is not wrong (we simply end up finding the eigenvalue of a $1\times1$
matrix, which is of course its only entry).  Some algebra reveals that the above effective Hamiltonian can be
rewritten in the following sensible form
\begin{equation}
\label{Heff2}
\mathcal{H}_{\mathrm{eff}}=\mathcal{J}^2c_1(\mathcal{V},\mathcal{M},n)\rho-\mathcal{J}^2c_2(\mathcal{V},\mathcal{M},n)\bm{\rho}\cdot\bm{S},
\end{equation}
where $\bm{S}=\bm{S}_a+\bm{S}_b$, the coefficients $c_1$ and $c_2$ are
presented in Appendix \ref{highfillinglobes}, and the vector superfluid
density $\bm{\rho}$ acts as a magnetic field.  It is not hard to show
that $c_2$ is strictly positive \footnote{For a given $n>1$, the Mott
  lobe is bounded below by $\mathcal{V}+n-1$ and above by
  $\mathcal{V}+n$, and it is within these constraints that the
  coefficient $c_2$ is strictly positive.}, and as a
result minimization of $\mathcal{H}_{\mathrm{eff}}$ is achieved
when $\bm{S}$ points along the magnetic field.  This result makes
perfect sense, since it amounts to self consistency in the
direction of the order parameter.  Because the quantum number of
total spin in the unperturbed groundstate is $s=(n-1)/2$, we find
\begin{equation}
\mathcal{B}(n)=\mathcal{J}^2\left(\frac{1}{\mathcal{J}}+c_1(\mathcal{V},\mathcal{M},n)-c_2(\mathcal{V},\mathcal{M},n)\frac{n-1}{2}\right).
\end{equation}
$\mathcal{B}(n)=0$ determines the boundary of the $n$-filling Mott
lobe, which we will denote by
$\mathcal{M}^{\mathrm{KH}}_n(\mathcal{J},\mathcal{V})$.  The first
three lobes are shown in Fig. \ref{JVphasediagram}(a).  Notice that the
width of the $n=1$ MI is increased (from $1$ to $1+\mathcal{V}$ along
the $\mathcal{M}$ axis), while the widths of the higher filling lobes
are unaffected.  The reason for this goes back to the discussion in
Sec. \ref{effectivespinmodelsSC}, where we pointed out that the charge
gap is given by $\Delta_{\mathrm{c}}=U(1+\mathcal{V})$ for the $n=1$
MI and by $\Delta_{\mathrm{c}}=U$ for an $n\geq2$ MI.

At this point we are in a position to see that something is wrong with
the phase boundaries as they have been presented so far [the solid red
lines in Fig. \ref{JVphasediagram}(a)].  Explicit calculation yields
the boundaries at zero Kondo coupling
\begin{equation}
\label{wronganswer}
2\mathcal{M}_n^{\mathrm{KH}}(\mathcal{J},0)=2n-1-\mathcal{J}\pm\sqrt{(\mathcal{J}-1)^2-4\mathcal{J}n-\frac{4J}{n+1}}
\end{equation}
whereas the mean-field phase boundaries for the single-band Bose Hubbard
model are given by
\begin{equation}
2\mathcal{M}_n^{\mathrm{H}}(\mathcal{J})=2n-1-\mathcal{J}\pm\sqrt{(\mathcal{J}-1)^2-4\mathcal{J}n},
\end{equation}
which only agree at $\mathcal{J}=0$.  But the conduction bosons of the
Kondo-Hubbard model certainly \emph{are} governed by the single band
Bose Hubbard model at $\mathcal{V}=0$, since they don't talk to the localized
spins.  What has gone wrong?  The problem is that
Eq. (\ref{wronganswer}) was derived by assuming that the mean-field
ground state in the MI phase has $s=(n-1)/2$, which is
true whenever $\mathcal{V}>0$.  However, there are also the excited
spin states with $s=(n+1)/2$, separated from the ground state manifold by $\Delta_n=\mathcal{V}(n+1)$, which have been ignored.  Formally, such a procedure is
correct for any finite $\mathcal{V}$, so long as
$\mathcal{J}\psi\ll\Delta_n$ is satisfied.  As $\mathcal{V}\rightarrow0$, the range
of validity shrinks, until eventually at $\mathcal{V}=0$ the
perturbative results break down for any finite $\psi$.

\begin{figure*}[!t]
\centering
\subfiguretopcaptrue

\subfigure[][]{
\includegraphics[width=5.45 cm]{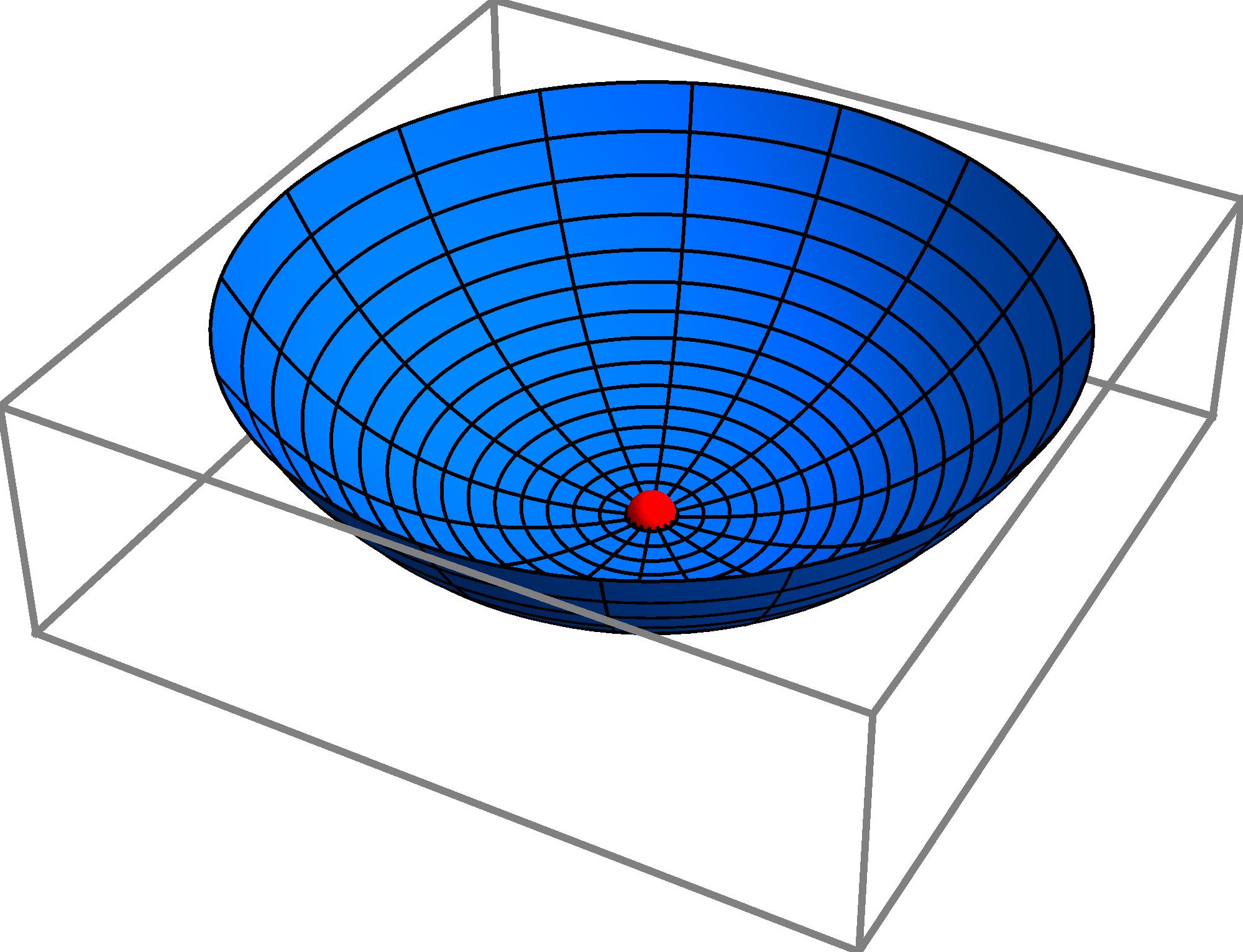}
\label{mh1}}
\subfigure[][]{
\includegraphics[width=5.45 cm]{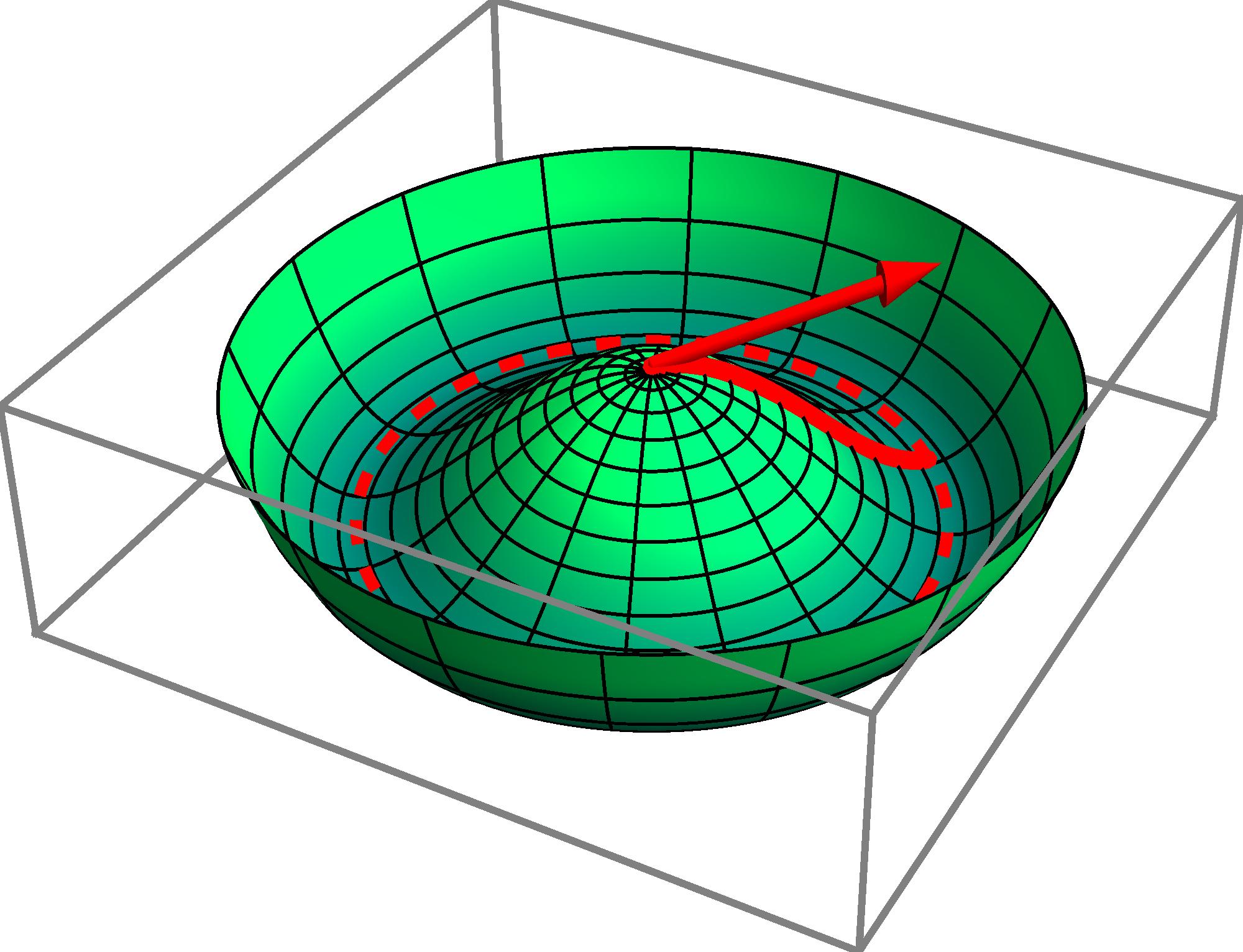}
\label{mh2}}
\subfigure[][]{
\includegraphics[width=5.45 cm]{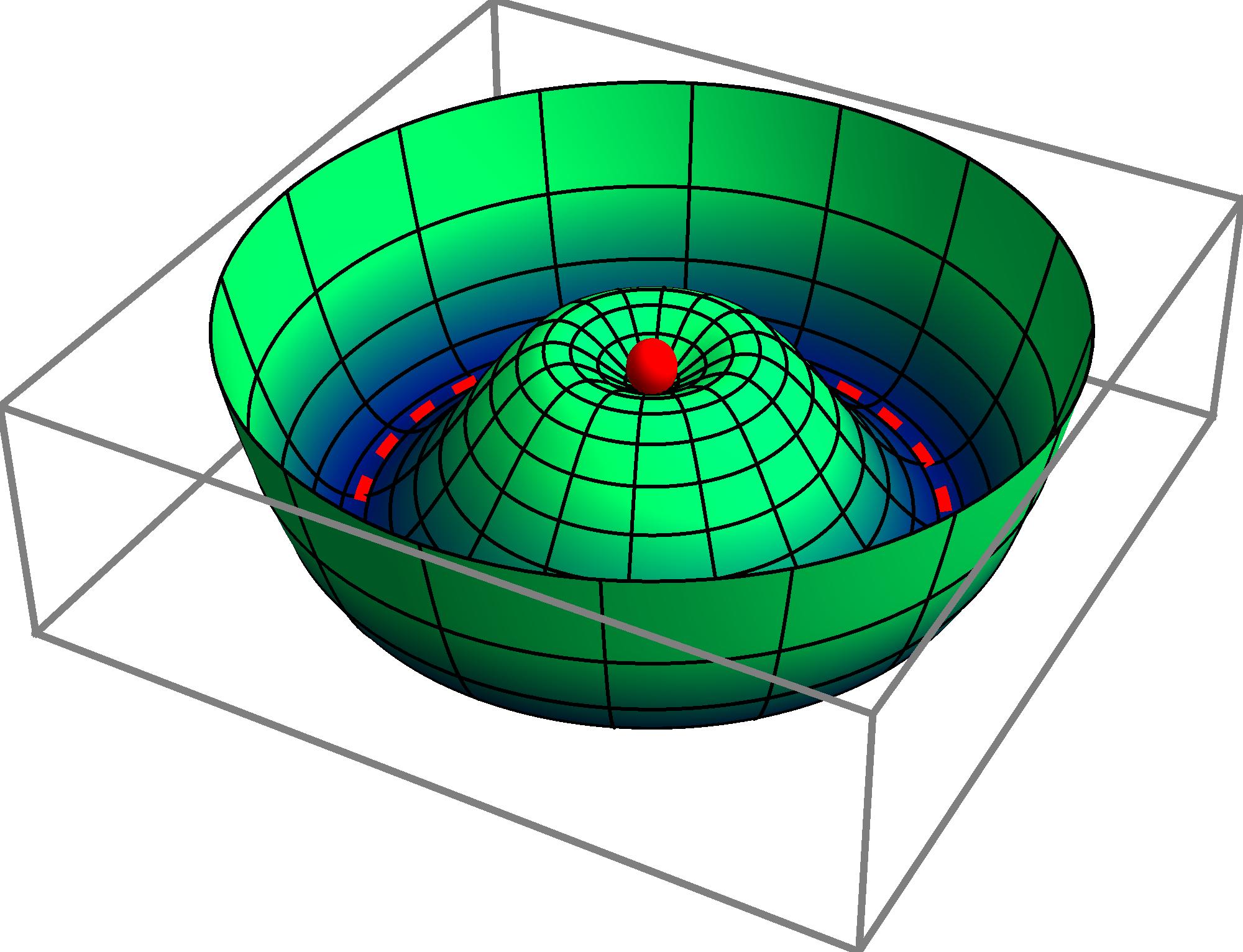}
\label{mh3}}

\caption{(Color online) Schematic illustration of the mean-field energy,
  demonstrating the mechanism that drives the SF to MI phase transition
  first-order.  \subref{mh1} and \subref{mh2} show the two forms that
  $E_{\mathrm{GA}}$ can take when $\mathcal{V}=0$, \subref{mh1} in the
MI phase and \subref{mh2} in the SF phase, with the order parameter
spontaneously breaking SU(2) symmetry by
choosing a particular direction.  In \subref{mh3} we show a possible
scenario for a small nonzero $\mathcal{V}$.  The superfluid is weakly affected (being rigid against spin
fluctuations), but when the order parameter is small magnetic fluctuations
are allowed, reducing the energy and causing the formation of a
metastable MI phase.}
\label{mexicanhat}
\end{figure*}

Once this issue is understood, it becomes clear that the phase transition must in fact be
first-order, and the argument is as follows. If we sit somewhere in
the red shaded region of Fig. \ref{JVphasediagram}(a), at $\mathcal{V}=0$ the
system is a superfluid (since we are outside of the dashed blue line,
which \emph{is} correct at $\mathcal{V}=0$), and so $E_{\mathrm{GA}}$
looks like it does in Fig. \ref{mh2}.  Now as we turn on small but finite
$\mathcal{V}$, a local minimum of the energy \emph{must} immediately
develop at $\psi=0$ [see Fig. \ref{mh3}], since perturbation theory has a finite range of
validity in $\psi$ and predicts $\mathcal{B}>0$ [a Mott insulator, see
the red dotted curve in Fig. \ref{JVphasediagram}(a)].  As $\mathcal{V}$
becomes larger, the range of validity for the perturbative
result Eq. (\ref{wronganswer}) increases, until eventually the $\psi=0$ minimum must be the global
minimum; at this point there is a first-order transition into a Mott
insulator of Kondo singlets.  Notice that the first-order transition is not tied to the development
of a cubic term in the energy, as is often the case, but rather
results from the negative contributions of even powers of $\psi$ beyond $\psi^2$.
By numerically solving the mean-field Hamiltonian we can obtain
the first-order phase boundaries, defined by where the metastable MI
and the superfluid become degenerate.  The boundary for the first Mott
lobe is shown in Fig. \ref{JVphasediagram}(b), whereas the
discontinuous jump in the order parameter across this boundary is shown in Fig. \ref{JVphasediagram}(c).

While the above argument is quite rigorous, it offers little physical
insight into what is really going on; there is a more intuitive
argument that explains the existence of the metastable MI phase and pinpoints the general features of our
model that give rise to it.  The superfluid ground state has, in a
sense to be made precise shortly, a certain rigidity against spin
fluctuations.  Denoting the $\mathcal{V}=0$ mean-field ground state
$|\langle\bm{S}_b\rangle,\bm{\rho}\rangle$, and letting
$\hat{\bm{\rho}}=\bm{\rho}/\psi^2$ be a unit vector in the
direction of the superfluid, one might guess that turning on a small $\mathcal{V}$
causes the groundstate to be the singlet-like
$|-\frac{1}{2}\hat{\bm{\rho}},\bm{\rho}\rangle-|\frac{1}{2}\hat{\bm{\rho}},-\bm{\rho}\rangle$,
gaining $3\mathcal{V}/2$ from the Kondo term.
However, it can be seen that such a state actually has
$\sim\mathcal{J}\psi^2$ more kinetic energy than the mean-field
ground state at $\mathcal{V}=0$.  Hence when $\mathcal{J}\psi^2\gtrsim\mathcal{V}$, rather than fluctuating its
spin the superfluid will just anti-align with the localized spin,
gaining only $\mathcal{V}/2$ from the Kondo term.  Notice that this
rigidity is purely kinetic in origin, and that the mean-field ground state
for small $\mathcal{V}$ agrees, both in form and energy, with the lowest order results of
Sec. \ref{effectivespinmodelsSC}.  On the other hand, when
$\mathcal{J}\psi^2\lesssim\mathcal{V}$, the superfluid gains more energy from
the Kondo term by fluctuating its spin than it looses in kinetic
energy, and hence its energy decreases by $3\mathcal{V}/2$.  The tendency
of the mean-field energy to be lowered more for small $\psi$ than
for large $\psi$ is what enables a local minimum at $\psi=0$ to arise
\footnote{It is also essential that for a superfluid density $\psi^2$,
  the depth of the superfluid minimum can become much smaller than
$\mathcal{J}\psi^2$ as $\mathcal{B}\rightarrow0$ (i.e., as the Mott transition
is approached).  If this were not the case, the larger energy gain
from the Kondo term at small $\psi$ might not
induce a local minimum at $\psi=0$.}.

\section{Experimental details\label{experimentaldetails}}
\subsection{The optical lattice\label{opticallattice}}
\begin{figure}[!t]
\centering
\includegraphics[width=8.0 cm]{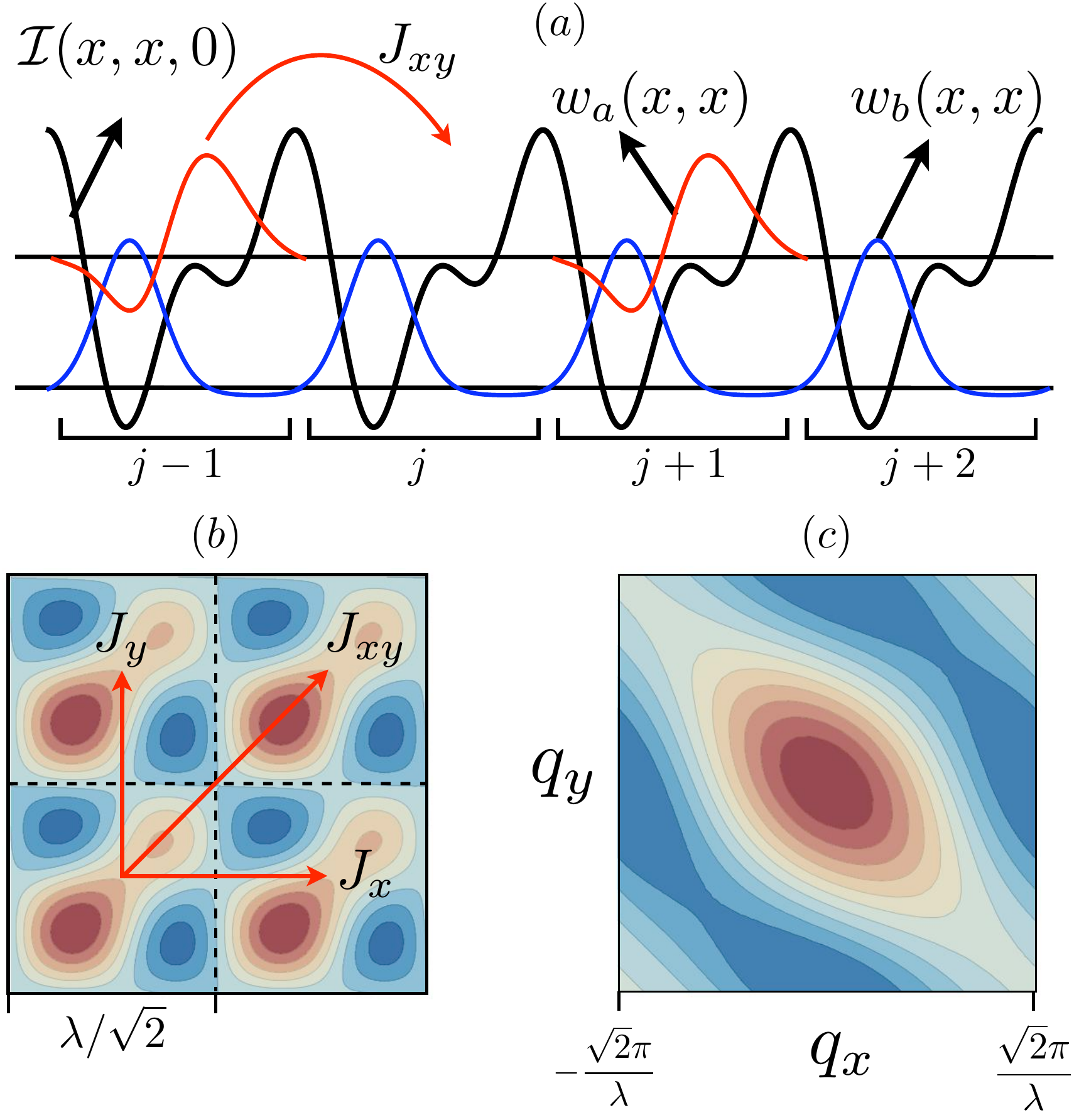}
\caption{(Color online). (a) Plot of $\mathcal{I}(x,x,0)$, with the wave functions $w_a$ and $w_b$ shown schematically.  (b) A
contour plot of $\mathcal{I}(x,y,0)$, showing the primary hoppings.  The resulting
spectrum in the $a$ band (c) can be fit to a tight-binding model with
hoppings $J_x$, $J_y$, and $J_{xy}$ [see (b)].}
\label{JVPlattice}
\end{figure}

As a specific example of how the approximations and parameters
considered in this manuscript can be obtained in an optical lattice, we
consider the 2D potential of Ref. \cite{sebbystrabley}
\begin{eqnarray}
\mathcal{I}(x,y)&=&-4\mathcal{I}_o[\cos{kx}+\cos{ky}]^2\nonumber \\
&-&\mathcal{I}_i[2\cos(2kx-2\varphi)+2\cos{2ky}],
\end{eqnarray}
and a deep transverse confining lattice with potential
$\mathcal{I}_{\perp}\cos(k_zz)$.  Each unit cell consists of a biased
double well [Fig. \ref{JVPlattice}(a)],
and control of the bias allows for an adjustment of the overlap integral
$\int dx dy |w_a(x,y)|^2|w_b(x,y)|^2$ and hence of $V$.  This lattice offers a large parameter space to play with, and here we just give one
example of a lattice configuration that could facilitate our model.
Defining $E_R=\frac{h^2}{2m\lambda^2}$, we choose the parameters
$\{\mathcal{I}_o,\mathcal{I}_i,\varphi\}=\{0.9E_R,2.52
E_R,0.3\pi\}$, for which the deep well has a depth of $\sim21E_R$ and
the shallow well has a depth of $\sim11E_R$.

Using a transverse lattice with $\mathcal{I}_{\perp}=40E_R$ and
$\lambda=2\pi k_z^{-1}=780\mathrm{nm}$, and the scattering length of $^{87}\mathrm{Rb}$, we find
$U_{b}/4J_b\approx2\times10^3$, $U_{a}/4J\approx10$, and
$V/U_a\approx0.1$.  Here $J_b$ is the nearest neighbor hopping
matrix element for the $b$ band.  This is just inside the $n=1$ mean-field Mott lobe, with
$V/U_a$ in the correct range to observe the first-order phase transition, and these parameters can be
adjusted to exit the Mott insulator in the $a$ band, or to change
$V/U_a$.  One consequence of this geometry is that the bands are not
isotropic.  The effect is very small for the $b$ band, but for the $a$
band, by fitting the numerically calculated dispersion [Fig. \ref{JVPlattice}(c)] to a tight binding model, we find
primary hoppings $J_x=J_y=0.002E_R$ and $J_{xy}=0.0035E_R$ [Fig. \ref{JVPlattice}(b)].

Other important energy scales are the band gaps, and for the
parameters given above the first 4 gaps (all those between the $b$
band and the $a$ band and the gap above the $a$ band) are all at least
$1E_R$.  This should be compared to the largest relevant interaction
energy $U_b$, and for the parameters given above we find
$U_b\approx0.28E_R$.  It should be noted that this comparison is an
extremely conservative metric of how the interaction energies compare to the band
gaps.  After all, $U_b$ is the largest interaction energy in the
model, and the band gap separating the $b$ band from the band directly
above it is larger than $5E_R$.

\subsection{Observation of the first-order transition\label{detection}}
In order to probe the SF to MI phase transition, an ideal starting point would be a 2D ($x-y$ plane) MI
of spin triplet pairs in the lowest vibrational level of the 2D
lattice with $\mathcal{I}_i=0$.  The PM MI could then be achieved with high
fidelity by ramping up $\mathcal{I}_i$ to establish an array of double
wells in the $x-y$ plane, and then using either Raman pulses
\cite{mullerbloch} or the population swapping techniques of
Refs. \cite{anderlini,wirth} to populate the $a$ band.  Standard
time-of-flight imaging, combined with band-mapping techniques
\cite{anderlini} and spin
selective imaging, could be used to resolve both the superfluid coherence and the magnetic
ordering in either band.

\begin{figure}[!h]
\centering
\includegraphics[width=8.0 cm]{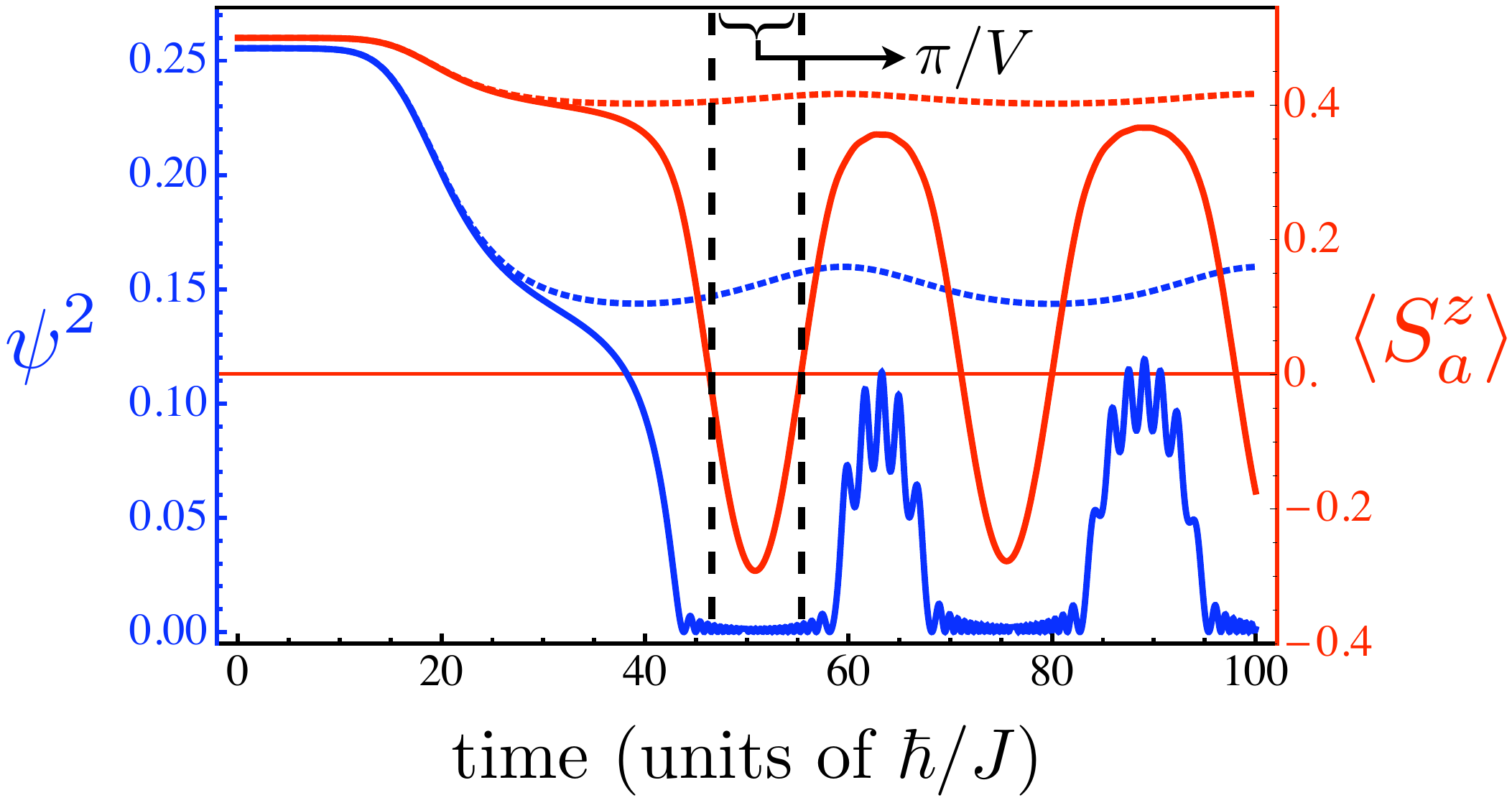}
\caption{(Color online) Dynamics during a slow ramp
of $\mathcal{V}$ from the SF to the MI at fixed conduction band filling $n=1$ (through a first-order phase
transition).  The ramping function is
$\mathcal{V}(t)=\mathcal{V}_f(\tanh(t/t_0)+1)/2$, with
$t_0\approx5.7\hbar/J$.  For a ramp to
$\mathcal{V}_f=\mathcal{V}_c+\delta~(\mathcal{V}_f=\mathcal{V}_c-\delta)$, the blue solid (dotted)
line is $\psi^2$ and the red solid (dotted) line is $\langle S^z_a\rangle$.  The fast oscillations of $\psi^2$ in
the non-adiabatic case occur on a time scale of order $\frac{2\pi}{U}$.
}
\label{Dynamics}
\end{figure}

As we have discussed in Sec. \ref{opticallattice}, $\mathcal{V}$ can be tuned in a double well
lattice.  Therefore the first-order phase transition should be observable by sitting just outside the
unit-filled Mott insulator lobe and increasing $\mathcal{V}$ from $0$ to some
value $\mathcal{V}_f$ large enough to support a Mott insulating ground state.  The
discontinuous nature of the phase transition will cause a failure of
adiabaticity for even an arbitrarily slow sweep of $\mathcal{V}$.
At the mean-field level, the effect of ramping $\mathcal {V}$ can be
explored with no further approximation by solving the time-dependent
equations of motion that result from minimization of the Lagrangian
\cite{jaksch}
\begin{equation}
\mathcal{L}=\langle i\frac{d}{dt}-\mathcal{H}_{\mathrm{GA}}\rangle,
\end{equation}
where the expectation value is taken in the single site Hilbert space
(for practical calculations, this space must be truncated by cutting
off the maximum number of allowed bosons).  Defining $\mathcal{V}_c$
to be the value of $\mathcal{V}$ at which the metastable SF solution
disappears (note that this is \emph{not} where we define the phase
boundary in Sec. \ref{meanfieldtheory}), we compare a slow ramp of $\mathcal {V}$ from $0$ to
$\mathcal{V}_f=\mathcal{V}_c\pm\delta$.  For $\mathcal{V}_f<\mathcal{V}_c$, we
observe a nearly adiabatic reduction of the SF component, whereas for
$\mathcal{V}>\mathcal{V}_c$ we observe collapses and
revivals of the SF component.  This is reminiscent of
the behavior seen in Refs. \cite{greinerbloch,will}, where a fast quench
was studied in the single-band Bose Hubbard model, but here the collapses occur
even for very slow lattice ramps.  Since the collapses are pinned to a rotation
of the magnetization [Fig. \ref{Dynamics}], it is not
surprising that they repeat on a time scale $\sim\frac{2\pi}{V}$.

Regarding the experimental feasibility of the proposed dynamics around the MI to SF
transition, the total time elapsed in Fig. \ref{Dynamics} is $~100\hbar/J$, which is comparable to the longest excited band decay times
measured in Ref. \cite{mullerbloch} for the $n=1$ situation.  This
suggests that dynamical evidence of the first-order transition, e.g. loss of
adiabaticity or hysteresis, should be within reach of current
experiments.  We also note that the first-order phase transition could
be explored by measuring local and static observables in the trap, for
instance it could manifest as a discontinuity in the density profile.

\section{Summary and conclusions\label{conclusion}}

In real metals, it is known that when conduction
electrons interact with magnetic impurities their behavior is
drastically altered; in order to describe such systems, it is
necessary to study many-body Hamiltonians that include spin, charge,
and orbital degrees of freedom, such as the KLM.  The bosonic
analogues of such systems are experimentally accessible using ultracold alkali
atoms in \emph{non-separable} optical lattices, motivating this fairly in depth study of
the Bose Kondo-Hubbard model.

Our primary new finding is that the SF to MI phase transition of the conduction
bosons is qualitatively modified by the existence of an arbitrarily
small Kondo coupling to a band of localized spins.  When approaching
the MI, as the superfluid density of the conduction bosons is reduced, magnetic fluctuations driven by the Kondo exchange
induce a metastable MI of Kondo singlets, and as a result the
associated phase transition becomes first-order.  We expect the first-order phase
transition to be observable in experiment, making this model a leading
candidate for observing entirely new, strongly correlated, multi-band
phenomena in optical lattices.

\section*{Acknowledgments}
We would like to thank S. F\"{o}lling, I. Bloch and L.-M. Duan for helpful discussions.  This work was supported by
grants from the NSF (PFC and Grant No. PIF-0904017), the AFOSR, and a grant from the ARO
with funding from the DARPA-OLE.

\appendix
\section{Derivation of the weak coupling Hamiltonian \label{wdderivation}}
Here we derive the second order weak-coupling effective Hamiltonian via
perturbation theory in the Kondo term $\mathcal{H}_V=2V\!\sum_j\bm{S}_{ja}\cdot\bm{S}_{jb}$
\begin{equation}
\mathcal{H}^{(2)}_{\mathrm{wc}}=\sum_{n,\Sigma}\frac{\mathcal{H}_{V}|n,\Sigma\rangle\langle
n,\Sigma|\mathcal{H}_{V}}{\varepsilon_0-\varepsilon_n}.
\end{equation}
The state $|n,\Sigma\rangle$ above has one conduction boson in the single
particle state $\psi_n(j)$, the rest of the conduction bosons in
$\psi_{0}(j)$, and a spin configuration labeled by the index $\Sigma$;
only states of this form contribute at this order.  Because the energy denominators have no dependence on $\Sigma$, the
sum over $\Sigma$ is a completeness identity in spin space and can be
dropped.  Using the basis transformation
$a^{\dagger}_{j\sigma}=\sum_{m}a^{\dagger}_{m\sigma}\psi_m(j)$ we can
rewrite the Kondo coupling as
\begin{equation}
\mathcal{H}_V=2V\sum_{jmn}\bm{S}^b_{j}\cdot\bm{S}^a_{mn}\psi_m(j)\psi_n(j),
\end{equation}
where $\bm{S}^a_{mn}=\sum_{\sigma\sigma'}a^{\dagger}_{m\sigma}\bm{\tau}_{\sigma\sigma'}a^{}_{n\sigma'}$.
For expectation values within the degenerate ground state manifold we
have the equivalence
\begin{equation}
\mathcal{H}^{(2)}_{\mathrm{wc}}=4V^2\sum_{mjl}\frac{\left(\bm{S}^b_j\cdot\bm{S}^a_{0m}\right)\left(\bm{S}^b_{l}\cdot\bm{S}^a_{m0}\right)\mathcal{G}^0_{jl}\mathcal{G}^{m}_{jl}}{\varepsilon_0-\varepsilon_m}.
\end{equation}
Taking advantage of common identities for pauli matrices, within the
ground state manifold we can rewrite
\begin{equation}
\left(\bm{S}^b_j\cdot\bm{S}^a_{0m}\right)\left(\bm{S}^b_{l}\cdot\bm{S}^a_{m0}\right)=\frac{\mathcal{N}_a}{4}\bm{S}^b_{j}\cdot\bm{S}^b_{l}-\frac{\delta_{jl}}{2}\bm{S}^a_{00}\cdot\bm{S}^b_{j},
\end{equation}
which leads to
\begin{eqnarray}
\mathcal{H}_{\mathrm{wc}}^{(2)}&=&-n\mathcal{N} V^2\sum_{j,l}\mathcal{R}_{jl}\bm{S}_{jb}\cdot\bm{S}_{lb} \nonumber \\
&+& 2V^2\bm{S}_a\cdot\sum_{j}\mathcal{R}_{jj}\bm{S}_{jb}.
\end{eqnarray}

\section{Parameters for $\mathcal{H}_{\mathrm{eff}}$.\label{highfillinglobes}}
The effective Hamiltonian that yields the Mott lobe boundaries
[Eq. \ref{Heff2}] follows from tedious but straightforward
manipulation of Eq. \ref{Heff1}.  The coefficients $c_1$ and $c_2$ are
found to be
\begin{eqnarray}
c_1=\frac{1}{2n}\left(\frac{n^2+2n}{\mathcal{M}-\mathcal{V}-n}-\frac{n^2-1}{1+\mathcal{M}-\mathcal{V}-n}\right.\nonumber\\
-\left.\frac{1}{1+\mathcal{M}-\mathcal{V}+\mathcal{V}n-n}\right)
\end{eqnarray}
and
\begin{eqnarray}
c_2=\frac{-1}{n^2+n}\left(\frac{n^2+2n}{\mathcal{M}-\mathcal{V}-n}-\frac{(n+1)^2}{1+\mathcal{M}-\mathcal{V}-n}\right.\nonumber\\
+\left.\frac{1}{1+\mathcal{M}-\mathcal{V}+\mathcal{V}n-n}\right).
\end{eqnarray}

\end{document}